\documentclass[aps,prd,twocolumn,showpacs,groupedaddress]{revtex4}
\usepackage{graphicx}
\usepackage{color}

\def\apj #1 #2 #3 {#1, ApJ, {\bf #2}, #3}
\def\apjl #1 #2 #3 {#1, ApJ, {\bf #2}, L#3}
\def\apjs #1 #2 #3 {#1, ApJS, {\bf #2}, #3}
\def\aap  #1 #2 #3 {#1, A\&A, {\bf #2}, #3}
\def\mnras #1 #2 #3 {#1, MNRAS, {\bf #2}, #3}
\def\pra #1 #2 #3 {#1, Phys.~Rev.~A., {\bf #2}, #3}
\def\prb #1 #2 #3 {#1, Phys.~Rev.~B., {\bf #2}, #3}
\def\prc #1 #2 #3 {#1, Phys.~Rev.~C., {\bf #2}, #3}
\def\prd #1 #2 #3 {#1, Phys.~Rev.~D., {\bf #2}, #3}
\def\pre #1 #2 #3 {#1, Phys.~Rev.~E., {\bf #2}, #3}
\def\prl #1 #2 #3 {#1, Phys.~Rev.~Lett., {\bf #2}, #3}
\def\plb #1 #2 #3 {#1, Phys.~Lett.~B., {\bf #2}, #3}
\def\science #1 #2 #3 {#1, Science., {\bf #2}, #3}
\def\nature #1 #2 #3 {#1, Nature., {\bf #2}, #3}
\def\nphysa #1 #2 #3 {#1, Nucl.~Phys.~A., {\bf #2}, #3}
\def\nphysb #1 #2 #3 {#1, Nucl.~Phys.~B., {\bf #2}, #3}
\def\nphysbs #1 #2 #3 {#1, Nucl.~Phys.~B.~Suppl., {\bf #2}, #3}

\def\h#1{\hbox{${}^{#1}$H}}

\def\h502{\hbox{$ h^{2}_{50}$}}

\def\fun#1#2{\lower3.6pt\vbox{\baselineskip0pt\lineskip.9pt
  \ialign{$\mathsurround=0pt#1\hfil##\hfil$\crcr#2\crcr\sim\crcr}}}
\begin{document}
\bigskip
\bigskip

\title{Constraints on preinflation fluctuations in a nearly flat open $\Lambda$CDM cosmology}
\author{
G. J. Mathews,$^{1,2}$ I.-S. Suh,$^1$ N. Q. Lan,$^{3}$ and T. Kajino$^{2,4}$
}
\address{$^1$University of Notre Dame, Center for Astrophysics, Notre Dame, IN 46556 USA
}
\address{$^2$National Astronomical Observatory of Japan, 2-21-1
Osawa, Mitaka, Tokyo, 181-8588, Japan
}
\address{$^3$Department of Physics, Hanoi National University of Education, 136 Xuan Thuy, Cau Giay, Hanoi 100000, Vietnam
}
\address{$^4$Department of Astronomy, Graduate School of Science, The
University of Tokyo, 7-3-1 Hongo, Bunkyo-ku, Tokyo, 113-0033, Japan
}
%


\date{\today}
\begin{abstract}
We analyze constraints on parameters characterizing the preinflating universe  in an open inflation model with a
present slightly open $\Lambda$CDM universe.  We employ an analytic model to show that for a broad class of inflation-generating effective potentials, the simple requirement that some fraction of the observed dipole moment represents a pre-inflation isocurvature fluctuation allows one to set upper and lower limits on  the magnitude and wavelength scale of preinflation  fluctuations in the inflaton field, and
the curvature of the preinflation universe, as a function of the fraction of the total initial energy density in the inflaton field as inflation begins. We estimate that if the preinflation contribution to the current Cosmic Microwave Background (CMB) dipole  is near the upper limit set by the {\it Planck} Collaboration then the current constraints on $\Lambda$CDM cosmological parameters allow for the possibility of a significantly open $\Omega_{i} \le 0.4$ preinflating universe for a broad range of the fraction of the total energy in the inflaton field at the onset of inflation.
This limit to $\Omega_{i}$ is even smaller if a larger dark-flow tilt is allowed.
\end{abstract}

\maketitle

\section{Introduction}
There is now a general consensus that cosmological observations have established that we live in a nearly flat universe.  The best fit of the combined CMB  + HiL +BAO fit by  the Planck collaboration \cite{Planck} obtained a closure content of the universe to be $\Omega_{0} = 1.005^{+0.0062}_{-0.0065}$ implying a curvature content of $\Omega_k \equiv  1 - \Omega_0 = -0.0005^{+0.0065}_{-0.0062}$.  Similarly, the  WMAP 9yr \cite{WMAP} analysis  obtained $\Omega_{0} = 1.0027^{+0.0038}_{-0.0039}$, or $\Omega_k = -0.0027^{+0.0039}_{-0.0038}$.  This is indeed very close to exactly flatness and there is a strong theoretical motivation to expect the present universe to be perfectly flat as a result of cosmic inflation.

Nevertheless, in this paper we consider the possibility that the present universe is slightly open, i.e. both CMB analyses allow  $\Omega_0 \ge 0.994$ at  the 95\% confidence level.  That being the case, then one can entertain the possibility that we are in a slightly open $\Lambda$CDM universe.  Indeed, it is well known that a matter-dominated  universe must eventually deviate from perfect flatness since $\Omega(t) = 1 - k/ [a(t)^2 H(t)^2]$ and the denominator eventually becomes small.  In a $\Lambda$CDM cosmology, however, as  the universe becomes cosmological-constant dominated,  then  $H(t) \rightarrow $ constant, and $a(t)$  grows exponentially, so that  flatness is eventually guaranteed.  However, since we have only recently entered the dark-energy epoch, there is still a possibility for a slight deviation  of $\Omega_0$ from unity.  In this case any curvature that existed before inflation might now be visible on the horizon.

In this paper, therefore, we explore the possibility that the  universe is slightly open with $\Omega_0 \approx 0.994$.  In this case a glimpse of preinflation fluctuations  could  just now be entering the horizon before the universe becomes totally dark-energy dominated and flat.  Our goal, therefore,  is to determine what constraints might  be placed on inhomogeneities  and curvature content in the preinflation universe based upon current cosmological observations.

There are  many possible paradigms for inflation in an open universe.  Most involve models \cite{Liddle00} in which there are two inflationary epochs.
For  open inflation models  \cite{Sasaki93,White14},  in the first epoch the universe must tunnel from a metastable vacuum state and then in a second epoch the universe slowly rolls down toward the true minimum.  In string landscape \cite{Yamauchi11}, for example,  such tunneling transitions to lower metastable vacua can occur through bubble nucleation.
Other possibilities include "extended open inflation"  \cite{Chiba00} in which a nominally coupled scalar
field with polynomial potentials exists for which there is a Coleman-de Luccia instanton, or that of two or multiple scalar fields \cite{Linde95a,LInde95b,LInde99, Sugimura12}, or a scalar-tensor theory in which a Brans-Dicke field has a potential along with a trapped scalar field  \cite{Green97}.  Of relevance to the present work is that such multiple field models of inflation allow for the existence of isocurvature fluctuations in the inflating universe.  That is a fluctuation in the energy density of the inflaton field is offset by a fluctuation in another field such that there is no net change in the curvature.  Isocurvature fluctuations are the main focus of this work for reasons described below.

We note that  preinflation fluctuations in the inflaton field could appear as  a cosmic dark flow \cite{Kashlinsky10, Kashlinsky11, Kashlinsky12} possibly detectable as a universal  cosmic dipole moment \cite{Mathews15}.
Indeed, if  a detection were made  it would be an exceedingly interesting as such apparent large scale motion could  be a remnant of the birth of the universe out of  the
M-theory landscape \cite{Mersini-Houghton09}, or  a remnant of multiple field inflation \cite{Turner91, Langlois96a,Langlois96b}.  Indeed, a recent analysis \cite{Chary15} of foreground cleaned {\it Planck} maps finds a small set of $\sim 2-4^o$ regions  showing a strong 143 GHz emission that could be interpreted as a preinflation residual fluctuation due to interaction with another universe in the multiverse landscape.   Of particular interest to the present work, however, is the possibility that a contribution to the large-scale CMB dipole moment could also be a remnant of  preinflation isocurvature fluctuations from any source, but  just visible on the horizon now \cite{Kurki-Suonio91} in a nearly flat present universe.

Previously, a detailed Baysian analysis  \cite{Valivita09} of constraints on isocurvature fluctuations and spatial curvature has  been made that place limits on the contribution of such fluctuations to the present CMB power spectrum.  Here, however, our goal is different.  We wish to examine constraints on the preinflation universe.  We utilize an analytic model originally developed in Ref.~\cite{Kurki-Suonio91} for an open cosmology with a planar inhomogeneity of wavelength less than the initial Hubble scale.  We update that model  for a $\Lambda$CDM cosmology and a broad class of inflation-generating potentials rather than the $\phi^4$ potential considered in that work.  In particular we generalize that model to consider isocurvature fluctuations.  We  show that for a broad class of inflation generating potentials, one has a possibility to utilize the limits on the dark-flow contribution to the CMB dipole (and higher moments) and current cosmological parameters to fix the amplitude and wavelength of isocurvature fluctuations as a function of energy content of the inflaton field as the  universe just entered the inflation epoch.

The possibility of scalar isocurvature fluctuations is not well motivated in the usual inflation paradigm.  However,  if more than one field contributes significantly to the energy density during inflation one can get isocurvature fluctuations.   In particular, it is well known \cite{Grishchuk78, Kashlinsky94,Langlois97} that for adiabatic fluctuations, even on the largest scales, a significant dipole contribution will also lead  to large power in the  quadrupole and higher multipoles.   Therefore, the fact that the observed quadrupole moment is 2 orders of magnitude smaller than the dipole moment implies that a significant fraction of the observed dipole could not be due to adiabatic fluctuations.  However, as we summarize below, it is possible \cite{Langlois97} to have a large dipole contribution to the CMB from a super-horizon isocurvature fluctuation without overproducing the observed  quadrupole and higher moments.

In this context, the recent  interest \cite{Kashlinsky10, Kashlinsky11, Kashlinsky12} and controversy \cite{Planckdf, Atrio-Barandela13} over  the prospect that the local observed dipole motion with respect to the microwave background frame may not be a local phenomenon but could extend to very large (Gpc) distances is particularly relevant.    This dark-flow (or tilt) is precisely how a preinflation  isocurvature inhomogeneity would appear as it begins to enter the horizon.

Attempts have been made \cite{Kashlinsky10,Kashlinsky11,Kashlinsky12}  to observationally detect such dark flow
by means of the kinetic Sunayev-Zeldovich (KSZ) effect.  This is a distortion
of the CMB background along the line of sight to a distant galaxy cluster due to the motion of the cluster with respect to the background CMB.  A detailed analysis of  the KSZ effect based upon the WMAP data \cite{WMAP} seemed to confirm that a dark flow exists out to at least 800 $h^{-1}$ Mpc \cite{Kashlinsky12}.  However, this was not  confirmed in a follow-up analysis using  the higher resolution data from the Planck Surveyor \cite{Planckdf}.  This has led to a controversy in the literature.  For example, it has been convincingly argued in  \cite{Atrio-Barandela13} that the background averaging method in the Planck Collaboration analysis may have  led to an obscuration of the effect.

Moreover, recent work \cite{Atrio-Barandela15}  reanalyzed the dark flow signal in the analysis WMAP 9 yr and the 1st yr Planck data releases using a catalog of 980 clusters  outside the Kp0 mask to remove the regions around the Galactic plane and to reduce the contamination due to foreground residuals as well as that of point sources.. They found a clear correlation between the dipole measured at  cluster locations in filtered maps proving that the dipole is indeed associated with clusters, and the dipole signal was dominated by the most massive clusters, with a statistical significance better than 99\%.  Their results are  consistent with the earlier analysis and imply the existence of a primordial CMB dipole of nonkinematic origin and a dark-flow velocity of $\sim 600 - 1,000$ km s$^{-1}$.

In another important analysis, Ma $et ~ al.$ \cite{Ma2011} performed  a Bayesian statistical analysis of the possible mismatch between the CMB defined rest frame and the matter rest frame.  Utilizing  various independent peculiar velocity catalogs, they found that  the magnitude of the velocity corresponding to the  tilt in the intrinsic CMB frame   was $\sim 400$ km s${^{-1}}$ in a  direction  consistent with previous analyses.   Moreover, for most catalogs analyzed, a vanishing dark-flow velocity was excluded at about the $2.5\sigma$ level.  Similar to the present work they also considered the possibility   that some fraction of the CMB dipole could  be intrinsic due to a large scale inhomogeneity  generated by preinflationary isocurvature fluctuations. Their conclusion that inflation must have persisted for 6 $e$-folds longer than that needed to solve the horizon problem is consistent with the constraints on the superhorizon preinflation fluctuations deduced in the present work.

Therefore, even though  the constraints set by the Planck Collaboration are consistent with no dark flow, a dark flow is still possible in their analysis \cite{Planckdf} up  to  a (95\% confidence level)  upper limit of 254 km s$^{-1}$.  This is also consistent with numerous attempts (e.g. summary in \cite{Mathews15}) to detect a bulk flow in the redshift distribution of galaxies.  Hence,  nearly half  of the observed CMB dipole could still correspond to  a cosmic dark-flow component.  We adopt this as a realistic constraint on the possible observed contribution of preinflation fluctuations to the CMB dipole. However, based upon the analyses in Refs.~\cite{Atrio-Barandela15, Ma2011} a dark flow as large as $\sim 1000$ km s$^{-1}$ is not yet ruled out.  Hence, we also, consider the constraints based upon this upper value for the dark-flow velocity.

\section{Model}
We consider isocurvature fluctuations in the scalar field of the preinflationary universe.  For simplicity, we assume that the fluctuations in the inflaton field will be offset by fluctuations in the radiation (or some other) field just before inflation, or that the decay of the inflaton field after it enters the horizon will produce CDM  isocurvature  fluctuations \cite{Valivita09}.
The energy density of a general inflaton field is
\begin{equation}
\rho_\phi = \frac{1}{2} \dot \phi^2 + \frac{1}{2a^2} (\nabla \phi)^2 + V(\phi)~~.
\label{rhophi}
\end{equation}
We will assume that the $\dot \phi^2/2$ term can dominate over  $V(\phi)$  initially, but eventually $V(\phi)$ will dominate as inflation commences.
The quantity  most affected initially by the
density perturbation in the scalar field is, therefore,   the kinetic $\dot \phi^2/2$ term as inflation begins.

We consider a broad range of general inflation-generating  potentials $V(\phi)$ to drive  inflation \cite{Liddle00} with the only restriction that they be continuously differentiable in
the inflaton field $\phi$, i.e. $dV/d\phi \ne 0$.   We also restrict ourselves to modest isocurvature fluctuations in the scalar field with a wavelength  less
than the initial Hubble scale.  This  allows one  to ignore the
gravitational reaction to the inhomogeneities.

Moreover, this allows one to describe the initial expansion
with fluctuations due to scalar-field perturbations on top the usual   LFRW metric characterized by a dimensionless scale factor $a(t)$.
\begin{equation}
ds^2 =  -dt^2 + a(t)^2 \biggl[\frac{ dr^2}{1 - k r^2} + r^2 d\Omega^2 \biggr]~~,
\label{LFRW}
\end{equation}
where we adopt the usual coordinates such that $k = -1$ for an open cosmology, and $a(t) = 1$ at the present time.

The particle horizon  is given by the radial null geodesic in these coordinates,
\begin{equation}
r_H(t) = a(t) \int_0^t \frac{dt'}{a(t')}~~.
\label{horizon}
\end{equation}
This is to be distinguished from the Hubble scale $H^{-1}$, which at any epoch is given by the Friedmann equation to be
\begin{equation}
\frac{1}{H(t) } = a(t)  \sqrt{ 1 - \Omega(t)}~~.
\label{otot}
\end{equation}

For small inhomogeneities, the coupled equations for the Friedmann equation and the  inflation can then be written
\begin{equation}
H^2 = \frac{8 \pi}{3 m_{Pl}^2} (\rho_r + \langle \rho_{\phi}\rangle ) + \frac{1}{a^2}~~,
\label{expansion}
\end{equation}
\begin{equation}
\ddot \phi = \frac{1}{a^2} \nabla^2 \phi - 3 H \dot \phi - V'(\phi)
\label{phiddot}
\end{equation}
where $H =H(t) =\dot a/a$ is the Hubble parameter, and  $\phi =\phi(t, x)$ is the inhomogeneous inflaton field in terms of comoving coordinate $x$.
The radiation energy density $\rho_r = \rho_{r,i}(a_i/a)^4$ with $\rho_{r,i}$ the initial mass-energy density in
the radiation field.   The brackets  $ \langle \rho_{\phi}\rangle$ denote  the average energy density in the inflaton field. That is, we decompose the energy density in the inflaton field into an average part and a fluctuating part.
\begin{equation}
\rho_\phi = \langle \rho_{\phi}\rangle  + \delta \rho_\phi~~.
\label{rhosplit}
\end{equation}

\subsection{Initial conditions}
We presume that the initial isocurvature  inhomogeneities are determined at or near the Planck time.
Hence,  we set the initial Hubble scale  equal to the Planck length,
\begin{equation}
H_i^{-1} = m_{Pl}^{-1}~~.
\label{hi}
\end{equation}
For simplicity, one can consider \cite{Kurki-Suonio91} plane-wave inhomogeneities in the inflaton field.
\begin{equation}
\phi(t,z) = \phi_i + \delta \phi_i \sin{\frac{2 \pi}{\lambda_i}(a_i z - t)}~~.
\label{deltaphi}
\end{equation}
The wavelength of the fluctuation can then be  parametrized \cite{Kurki-Suonio91} by ,
\begin{equation}
\lambda_i = l H_i^{-1} = \frac{l}{m_{Pl} }= l  \sqrt{1 - \Omega_i} a_i ~~,
\end{equation}
with $l$  dimensionless in the interval  $0 < l < 1$.

The energy density in the initial inflaton field, $\rho_{\phi,i}$  is constrained to be less than the Planck energy density.  From Eqs. (\ref{expansion}) and (\ref{hi}) this implies
\begin{equation}
 \rho_{\phi,i } \equiv f \Omega_i \frac{ 3 m_{Pl}^4}{8 \pi} ~~,~~ 0 < \Omega_i < 1 ~~, ~~ 0 < f < 1~~,
\label{const1}
\end{equation}
where $1-\Omega_i$ is the initial curvature in the preinflation universe, and $f$ is the fraction of the initial total energy density in the inflaton field.
If the largest inhomogeneous contribution is from the $\dot \phi^2/2$ term, then the amplitude of the inhomogeneity in Eq.~(\ref{deltaphi}) is similarly constrained to be
\begin{equation}
\frac{\delta \phi_i}{m_{Pl}} = \biggl(\frac{3 f \Omega_i l^2}{16 \pi^3}\biggr)^{1/2} ~~.
\end{equation}
Hence, the shorter the wavelength, the smaller the amplitude must be for the energy density not to exceed the Planck density. The maximum initial amplitude we consider is therefore
\begin{equation}
(3/16\pi^3)^{1/2}  m_{Pl}  =0.078 m_{Pl}~~,
\end{equation}
 for fluctuations initially of a Hubble length.

Hence, our assumption that one can treat the fluctuation as a perturbation on top of an average LFRW expansion is reasonable.
Fluctuations beyond the Hubble scale can of course have larger amplitudes, but those are not considered here.
Note also, however,  that the assumption of ignoring the effect of gravitational perturbations on the inflaton field  in Eq. (\ref{phiddot}) is justified as long as we restrict ourselves to fluctuations less than the initial Hubble scale  $H \lambda < 1$.

At the initial time $t_i$  we have $H_i \lambda_i \equiv l < 1$. After that   the comoving wavelength  $H \lambda$ decreases until inflation begins.  During inflation then $H \lambda$  increases until a time $t_x$ at which $H_x \lambda_x = 1$.   At this  time the fluctuation exits  the horizon and is frozen in until it reenters the horizon at the present time.  How much $H \lambda$ decreases during the time interval from $t_i$ to $t_x$ depends upon the initial closure parameter $\Omega_i$ \cite{Kurki-Suonio91}.

 \subsection{An analytic model}
The problem, therefore, has three cosmological  parameters, $\Omega_i$, $l$, and $f$,  plus  parameters related to the inflaton potential $V(\phi)$.
We now develop upon a simple analytic model \cite{Kurki-Suonio91} to show that the inflaton potential can be constrained \cite{Liddle00} from the COBE \cite{Smoot92} normalization of fluctuations in the CMB for any possible  differentiable inflaton potential. We will also show that  the initial wavelength parameter $l$ and the initial closure $\Omega_i$ can be constrained for a broad range of scalar-field energy-density contributions $f$ by two requirements.  One is  that the resultant dipole anisotropy does not exceed the currently observed upper limit \cite{Planckdf} to the contribution to the CMB dipole moment.  The other is that the  higher multipole components not contribute significantly to the observed CMB power spectrum.

To begin with, the equation of state for the total density  in Eq. (\ref{expansion}) can be approximated as
\begin{equation}
\rho_r + \langle \rho_{\phi}\rangle  \approx A \biggl(\frac{a_i}{a}\biggr )^4 + B~~,
\end{equation}
where $A = \rho_{r,i}$, and $B= (3 m_{pl}/8\pi))V(\phi_i)$ are constants.  Explicitly,  from $t_i $ to $t_x$,
we invoke the slow-roll approximation.
Another simplifying assumption is that $V(\phi) \sim B$ is initially small compared to the matter density for the first scale (the one we are interested) to cross the horizon.
This assumption was verified in  \cite{Kurki-Suonio91} by a numerical solution of the equations of motion.

With these assumptions, the solution \cite{Kurki-Suonio91} of Eq.~(\ref{expansion}) for the scale  factor  at horizon crossing is simply,
\begin{equation}
\biggl( \frac{a_x}{a_i} \biggr) = \biggl( \frac{1 - l^2 (1 - \Omega_i)}{B l^2} \biggr)^{1/2} ~~.
\label{ax}
\end{equation}
This analytic approximation was also verified  to be accurate to a few percent by detailed numerical simulations in \cite{Kurki-Suonio91}.

We are especially interested in the case where the
length scales of these fluctuations were not expanded by inflation to be to many orders of magnitude larger than the present observable scales.  That is,
we have  the minimal amount of inflation such that the preinflation horizon is just visible on the horizon now.

The energy density in the fluctuating part of  the  inflaton field given in Eqs.~(\ref{rhophi})  and (\ref{rhosplit}) can be written as
\begin{equation}
\delta \rho_\phi = \frac{1}{2} \delta (\dot \phi^2) + \frac{1}{2a^2} \delta (\nabla \phi)^2 + \delta \rho_r + \delta V~~,
\end{equation}
while the average  part of the total energy density  plus pressure can be written
\begin{equation}
\rho + p = \langle \dot \phi^2\rangle   + \frac{1}{3a^2} \langle (\nabla \phi)^2 \rangle  + \frac{4}{3}  \rho_r ~~.
\end{equation}
Ignoring the gradient term that decays away as $a^{-4}$  we can express the approximate amplitude when a fluctuation exits the horizon to be
\begin{equation}
\frac{\delta \rho}{\rho + p} \bigg |_x \approx \frac{(1/2) \delta(\dot \phi)^2_x + \delta V_x}{\dot \phi^2_x}~~.
\end{equation}
Now using the slow-roll condition
\begin{equation}
\dot \phi = \frac{V'(\phi)}{3 H}~~,
\end{equation}
and Eq. (\ref{ax}), this  reduces \cite{Kurki-Suonio91}  to
\begin{equation}
\frac{\delta \rho}{\rho + p} \bigg |_x \approx K \frac{\sqrt{f \Omega_i} l^2}{[1 - l^2(1 - \Omega_i)]^{3/2}}~~,
\label{delrho}
\end{equation}
where the constant $K$ is given by:
\begin{equation}
K = \biggl[ 1 + \frac{3}{2 \pi} \biggr] 8 \pi \sqrt{2} \frac{V(\phi_i)^{3/2}}{V'(\phi_i) m_{Pl}^3}~.
\label{keq}
\end{equation}
What remains is to fix the normalization of the inflaton potential in Eq.~(\ref{keq}).

\subsection{Normalization of inflaton potential}
The usual quantum generated adiabatic fluctuations  during inflation are produced  from the same inflaton potential and are responsible for the fluctuations observed in higher moments of the CMB power spectrum.   Hence,  the observed power spectrum of the higher moments of the CMB can be used to fix the parameters of the inflaton potential that enters in Eq.~(\ref{keq}).

In the slow-roll approximation, the amplitude of the inflation-generated quantum  fluctuations  $\delta \rho_Q$ as they exit the horizon at $t = t_x$ are \cite{Liddle00}
\begin{equation}
\delta_H \equiv \frac{\delta \rho_Q}{\rho + p} \bigg |_x = \biggl(\frac{4}{25}\biggr)^{1/2}\frac {H}{\dot \phi} \frac {H}{2 \pi}~~,
\end{equation}
and for these, one has \cite{Liddle00}
\begin{equation}
\delta_H \approx   \biggl[ \frac{1}{\sqrt{75} \pi }\biggr] \frac{V^{3/2}(\phi_x)}{m_{Pl}^3 V'(\phi_x)}~~.
\label{delh}
\end{equation}
The COBE \cite{Smoot92} normalization ($\delta_H = 1.91 \times 10^{-5}$) of the CMB power spectrum requires \cite{Liddle00} ,
\begin{equation}
\frac{V^{3/2}(\phi_x)}{m_{Pl}^3 V'(\phi_x)} = 5.20 \times 10^{-4}~~.
\end{equation}
Since the same inflation-generating potential has expanded the preinflation fluctuation of interest here, we can deduce the constant $K$, independently of the analytic form of the potential,
 \begin{equation}
K = 5.2 \times 10^{-4} \biggl[ 1 + \frac{3}{2 \pi} \biggr] 8 \pi \sqrt{2} ~ = ~ 0.0270.
\end{equation}

\subsection{CMB fluctuations}

Having deduced the magnitude of the fluctuations in the cosmic energy density due to the
appearance on the horizon of isocurvature fluctuations from the preinflation era, it still remains to determine the constraints based upon the observed
CMB dipole and higher moments.

In particular, it must be demonstrated that a large dipole moment can be invoked without a simultaneous large amplitude in higher moments \cite{Grishchuk78,Kashlinsky94}.
This  has been definitively addressed in Ref.~\cite{Langlois97} where it was  demonstrated that two criteria must be satisfied: 1) one must have isocurvature fluctuations beyond the current horizon; and 2) their spectrum must be suppressed on smaller scales.  To derive the constraints based upon the explicit model considered here, we summarize the isocurvature derivation for large scales here.

As usual, the CMB temperature fluctuations are decomposed into spherical harmonics:
\begin{equation}
\frac{\Delta T}{T}(\theta, \phi) = \sum_{l = 1}^\infty \sum_{m=-l}^l a_{l m} Y_{l m}(\theta, \phi)~~.
\end{equation}
For a random Gaussian field the coefficients of this expansion relate to the power spectrum in wave number $k$ via the window function $ {\cal W}_l(k)$,
\begin{equation}
\langle \vert a_{l m} \vert^2 \rangle = \frac{4 \pi}{9} \int \frac{dk}{k} {\cal P}_S(k) {\cal W}_l^2(k)~~,
\end{equation}
where the power spectrum for the  sinusoidal fluctuation  considered here is simply related to the amplitude given in Eq.~(\ref{delrho}),
i.e.
\begin{equation}
{\cal P}_S(k) = \frac{1}{8}\biggl[ \frac{1 }{3}  \frac{\delta \rho}{\rho + P}\biggr]^2 \delta(k- k_0)~~,
\end{equation}
where, $k_0$ is the present wave number corresponding to the preinflation fluctuation.  Here and in what follows we treat $k$ in units of the present hubble scale $H_0^{-1}$.

Once the power spectrum is given, the only dependence upon the various multipoles is given by the window functions ${\cal W}_l$
For the case of a flat $\Omega = 1$ cosmology, the window functions all vanish and there is no contribution from superhorizon fluctuations as expected.
However, for the case of a nearly flat open cosmology of interest here, one can expand \cite{Langlois97} the relevant window functions in terms of the parameter
\begin{equation}
\epsilon = \sqrt{1 - \Omega_0}
\end{equation}
To leading order in $\epsilon$,  and for largest scales ($k <<1$) the isocurvature window functions at the surface of last scattering ($z \approx 1300$)   can be reduced
\cite{Langlois97} to simple functions of $k$ and $\epsilon$.
The dipole window function becomes
\begin{equation}
{\cal W}_{1k} \approx \frac{10}{3} k \epsilon  + {\cal O}(\epsilon^3)~~,
\end{equation}
while, the quadrupole term becomes,
\begin{equation}
{\cal W}_{2k} \approx \frac{\sqrt{24}}{5 \sqrt{3}} k  \epsilon^2   + {\cal O}(\epsilon^4)~~,
\end{equation}
From this we deduce that the r.m.s. dipole moment corresponding to a fluctuation at $k_0 $ is
\begin{equation}
\langle \vert a_{1}\vert^2 \rangle^{1/2}  \approx \sqrt{\frac{\pi}{2}} \frac{10}{9} \biggl[ \frac{\delta \rho}{\rho + p}\biggr] \epsilon \sqrt{k_0}~~,
\label{a1}
\end{equation}
while the quadrupole moment becomes
\begin{equation}
\langle \vert a_{2}\vert^2 \rangle^{1/2}  \approx \frac{4}{3} \sqrt{\frac{2\pi}{3}} \biggl[ \frac{\delta \rho}{\rho + p}\biggr] \epsilon^2 k_0 ~~,
\end{equation}
where
\begin{equation}
\langle \vert a_{l}\vert^2 \rangle \equiv \sum_m \langle \vert a_{l m}  \vert^2 \rangle~~.
\end{equation}
Requiring that the ratio of the quadrupole (and higher moments) to the preinflation component be less than the ratio of observed moments  places a constraint on the scale of the fluctuation.  Specifically we have
\begin{equation}
k_0   < \frac{1}{1 - \Omega_0}\frac{25 }{48} \frac{\langle \vert a_{2} \vert^2 \rangle} {\langle \vert a_{1}\vert^2 \rangle}~~.
\end{equation}
The observed CMB temperature is $T = 2.7258 \pm 0.00057$ \cite{Fixen09}.  The magnitude of the dipole moment with respect to the frame of  the Local Group is 5.68 mK \cite{Kogut93}.  From this we obtain $\langle \vert a_{1}\vert^2 \rangle <  32.3 \times10^6~\mu$K$^2$.  However this represents the net sum of intrinsic tilt plus local proper motion.
If we assume that  the Planck Collaboration upper limit of $v_{DF} < 254$ km s$^{-1}$ \cite{Planckdf}  for a dark flow velocity is in the same direction as the proper motion.  This  implies a preinflation dipole moment tilt of  2.30 mK, or $\langle \vert a_{1}\vert^2 \rangle <  5.29 \times 10^6 ~\mu$K$^2$.  However if we adopt an upper limit to the dark flow velocity of $1000$ km s$^{-1}$ then the dark flow must be in an opposite direction to the local proper motion and this would correspond to 9.06 $m$K, or $\langle \vert a_{1}\vert^2 \rangle <  82.0 \times 10^6 ~\mu$K$^2$.

As is well known, the quadrupole moment is suppressed in the CMB power spectrum.  The WMAP 9yr data lists
\begin{equation}
C_2 \equiv \frac{1}{5} \sum_m \langle \vert a_{2 m} \vert^2 \rangle = 157~\mu{\rm K}^2
\end{equation}
from which we obtain $\langle \vert a_{2}\vert^2 \rangle = 785 ~\mu$K$^2$.  Therefore, for $1 - \Omega_0 = 6\times 10^{-3}$ we deduce
\begin{eqnarray}
k_0 &<&  0.013~~ (v_{DF} = 254~{\rm km~s^{-1}})  ~~\nonumber \\
&< &  8.3 \times 10^{-4}~~ (v_{DF} = 1,000~{\rm km~s^{-1}})~~.
\end{eqnarray}

In an open $\Lambda$CDM universe,
the largest observable scale, that of the cosmic microwave background, has the comoving size
\begin{equation}
\frac{r_l}{ a_0}   = \frac{1}{H_0} \int _0^1 \frac{dx}{\sqrt{\Omega_\Lambda x^4 + (1 - \Omega_0)x^2 + \Omega_m x + \Omega_\gamma}} ~~.
\end{equation}
For a nearly flat $\Omega_{0} = 0.994$ cosmology we can adopt values  $\Omega_\Lambda = 0.697$, and $\Omega_m = 0.297$, (with $\Omega_\gamma = 0$)  that are consistent with the Planck \cite{Planck}  and WMAP
\cite{WMAP} results.  For these parameters, then ${r_l}/{ a_0} \approx {3.3}/{H_0}$.
So $k_0 = 1$  would correspond to the present Hubble scale, and $k_0 = 0.3$ corresponds to the present horizon. Hence, in order for a preinflation fluctuation to contribute to  the dipole moment while not affecting the quadrupole and higher moments, the preinflation isocurvature fluctuation corresponding to a dark flow velocity of $254$ km s$^{-1}$ must reside at $> 0.013^{-1} \approx 77$ times the present Hubble scale, or 23 times the present horizon implying that inflation persisted for 3.1 $e$-folds beyond that required to solve the horizon problem.  On the other hand, if a dark flow tilt were  as large as 1000 km s$^{-1}$, then the preinflation fluctuation would reside at 361 times the current horizon implying that inflation persisted about 6 $e$-folds beyond that needed to satisfy the horizon problem.  This is similar to the conclusions in \cite{Langlois97, Ma2011}.

\subsection{Constraints on preinflation parameters}

From our deduced values for $k_0$ and adopted contributions to the CMB dipole, Eq.~(\ref{a1}), we can infer  the amplitude of the preinflation fluctuation, i.e.
\begin{eqnarray}
\biggl[ \frac{\delta \rho}{\rho + p}\biggr]  &>& 0.068~~ (v_{DF} = 254~{\rm km~s^{-1}})~~\nonumber \\
&> &  0.27~~ (v_{DF} = 1,000~{\rm km~s^{-1}})~~.
\end{eqnarray}

Then, from Eq.~(\ref{delrho}) we have a relation between the fraction of energy in the inflaton field and the other parameters,
\begin{equation}
f = K^2 \biggl[ \frac{\delta \rho}{\rho + p}\biggr] ^2 \frac{[1 - l^2(1 - \Omega_i)]^3}{\Omega_i l^4} < 1
\end{equation}

\begin{figure}[htb]
\includegraphics[width=3.5in,clip]{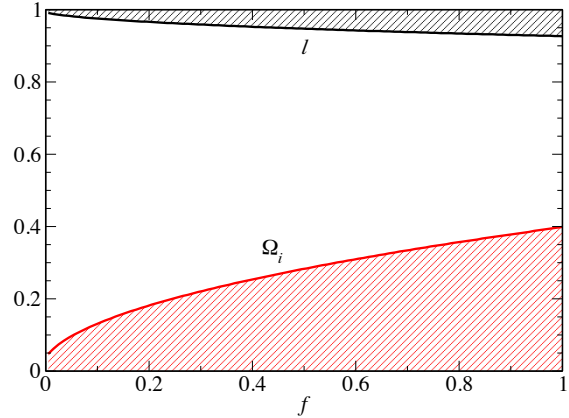}
\caption{ (Color online) Constraints on the preinflation parameters as a function of the fraction $f$ of the initial preinflation energy density in the inflaton field for a preinflation fluctuation corresponding to a present tilt velocity of 254 km s$^{-1}$ from the upper limit to the  {\it Planck} analysis \cite{Planckdf}.
Lower shaded region shows allowed values for the initial closure parameter $\Omega_i$.  Upper shaded region shows the allowed values of the wavelength parameter $l$ for  preinflation isocurvature fluctuations in the inflaton field.}
\label{fig:1}
\end{figure}

\begin{figure}[htb]
\includegraphics[width=3.5in,clip]{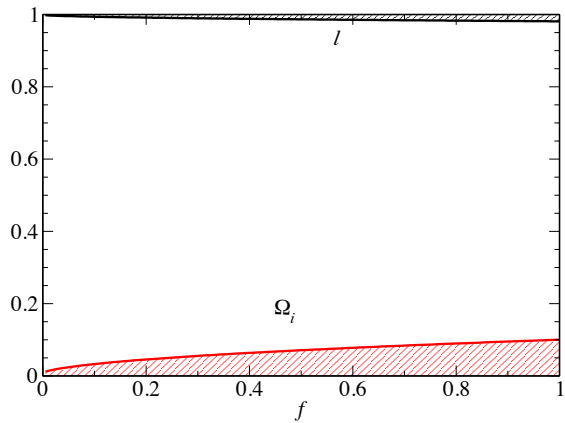}
\caption{(color online) Constraints on the preinflation parameters as a function of the fraction $f$ of the initial preinflation energy density in the inflaton field for a preinflation fluctuation corresponding to a present tilt velocity of 1000 km s$^{-1}$ from the upper limit of Ref.~\cite{Atrio-Barandela15}.
Lower shaded region shows allowed values for the initial closure parameter $\Omega_i$.  Upper shaded region shows the allowed values of the wavelength parameter $l$ for  preinflation isocurvature fluctuations in the inflaton field.}
\label{fig:2}
\end{figure}

Figure \ref{fig:1} summarizes values for $\Omega_i$ and $l$ that satisfy the constraint $f < 1$  based upon $v_{DF} = 254$ km s$^{-1}$  from the upper limit of the {\it Planck} analysis, and that the quadrupole and higher moments not exceed the value from the observed power spectrum.  Similarly, Figure \ref{fig:2} shows values for $\Omega_i$ and $l$ that satisfy the constraint $f < 1$  based upon $v_{DF} = 1000$ km s$^{-1}$.

The upper region in both figures shows that only values of $l $ near unity can satisfy this constraint while the upper limit to the initial closure parameter is $\Omega_i < 0.4$ ($\Omega_{k,i} > 0.6$) as $f \rightarrow 1$. This value reduces to $\Omega_i < 0.4$ if a dark flow of 1000 km s$^{-1}$ is allowed.


\section{Conclusion}

Although open inflation models are not particularly appealing because they imply that only enough inflation occurred such that the preinflation curvature is just beginning to enter the horizon, they are also interesting because they suggest that there may be an  opportunity to glimpse the preinflation universe. The current constraints on cosmological parameters from the Planck Collaboration \cite{Planck} and the WMAP analysis \cite{WMAP} suggest flatness, but are still  consistent with a slightly open ($\Omega_0 > 0.994$) nearly flat $\Lambda$CDM cosmology.

Here we have analyzed  a chaotic open inflationary universe  characterized by a general inflaton effective potential, but in which there is a plane-wave isocurvature fluctuation in the power spectrum.  We have shown in a simple analytic model that such fluctuations are constrained by the requirement that they not exceed the observed limit on the preinflation dipole contribution  deduced in the {\it Planck} analysis \cite{Planckdf} or the magnitude of the quadrupole and higher moments in the CMB power spectrum.  Indeed, from these constraints alone we find that the  preinflation fluctuation in the power spectrum must reside at least $\sim 80$ times the current Hubble scale.  Such fluctuations  are also constrained  by the near flatness of the current universe.  Indeed, all together we find that the wavelength of the preinflation fluctuation must be of order the Hubble scale as inflation begins.  Also, if there is a preinflation  component to the current cosmic dipole moment, then the initial preinflation closure parameter could have been as large as   $\Omega_i < 0.4$ ($\Omega_{k,i} > 0.6$).  This parameter reduces to $\Omega_i < 0.1$ if a dark flow as large as 1000 km s$^{-1}$ is allowed.

\begin{acknowledgments}
Work at the University of Notre Dame is supported by the U.S. Department of Energy under Nuclear Theory Grant DE-FG02-95-ER40934.
Work in Vietnam supported is supported in part by the Ministry of Education (MOE) Grant No. B2014-17-45.
Work at National Astronomical Observatory of Japan (NAOJ) was supported in part by Grants-in-Aid for Scientific Research of Japan Society for the Promotion of Science (JSPS)~ (26105517, 24340060).
\end{acknowledgments}


\begin{references}
%
\bibitem{Planck} Planck Collaboration, Planck XVI,  Astron. Astrophys. {\bf 566}, A54 (2014).
%
\bibitem{WMAP} G. Hinshaw  $et ~ al.$ ({\it WMAP Collaboration})    Astrophys. J. {\bf 208}, 19 (2013).
%
\bibitem{Liddle00} A. R. Liddle and D. H. Lyth,  {\it Cosmological Inflation and Large Scale Structure} (Cambridge University Press, Cambridge, England, 2000).
%
\bibitem{Sasaki93} M. Sasaki, T. Tanalka, K.Yamamoto, and J. Yokoyama, Phys. Lett. B {\bf 317}, 510 (1993).
%
\bibitem{White14} J. White, Y.-L Zhang and M. Sasaki, Phys. Rev. D {\bf 90} 083517 (2014).
%
\bibitem{Yamauchi11}  D. Yamauchi, A. Linde, A. Naruko, M. Sasaki and T. Tanaka, Phys. Rev. D {\bf 84}, 043513  (2011).
%
\bibitem{Chiba00} T. Chiba and M. Yamaguchi,  Phys. Rev. D {\bf 61}, 027304  (1999).
%
\bibitem{Linde95a} A. D. Linde, Phys. Lett. B {\bf 351}, 99 (1995).
%
\bibitem{LInde95b} A. D. Linde and A. Mezhlumian,  Phys. Rev. D {\bf 52}, 6789 (1995).
%
\bibitem{LInde99} A. Linde and M. Sasaki and T. Tanaka,  Phys. Rev. D {\bf 59}, 123522 (1999).
%
\bibitem{Sugimura12} K. Sugimura, D. Yamauchi and  M. Sasaki, J. Cosmol. Astropart. Phys. 01 ({\bf 2012}) 027.
%
\bibitem{Green97} A. M. Green and A. R. Liddle, Phys. Rev. D {\bf 55}, 609, (1997).
%
\bibitem{Kashlinsky10} A. Kashlinsky, F.  Atrio-Barandela, H.  Ebeling, A. Edge and D. Kocevski,  Astrophys. J. {\bf 712}, L81 2010.
%
\bibitem{Kashlinsky11} A. Kashlinsky, F. Atrio-Barandela, and H. Ebeling,  Astrophys. J. {\bf 732}, 1 (2011).
%
\bibitem{Kashlinsky12} A. Kashlinsky, F. Atrio-Barandela, and H. Ebeling,  arXiv:1202.0717 (2012).
%
\bibitem{Mathews15} G. J. Mathews, B. Rose, P. Garnavich, D. G. Yamazaki and T. Kajino, {\it to be published}.
%
\bibitem{Mersini-Houghton09} L. Mersini-Houghton and R. Holman, J. Cosmol. Astropart. Phys. 02, ({\bf 2009}) 6.
%
\bibitem{Turner91} M. S. Turner,  Phys. Rev. D {\bf 44}, 3737 (1991).
%
\bibitem{Langlois96a} D. Langlois, and T.  Piran,  Phys. Rev. D {\bf 53}, 2908 (1996).
%
\bibitem{Langlois96b} D. Langlois,  Phys. Rev. D {\bf 54}, 2447 (1996).
%
\bibitem{Chary15} R. Chary, arXiv:1510.00126v1.
%
\bibitem{Kurki-Suonio91} H. Kurki-Suonio, F. Graziani, and G. J. Mathews, Phys. Rev. D {\bf 44}, 3072 (1991).
%
\bibitem{Valivita09} J. Va\"liviita and T. Giannantonio, Phys. Rev. D {\bf 80}, 123516 (2009).
%
\bibitem{Grishchuk78} L. Grishchuk and Ya. B. Zel'dovich, Sov. Astron. {\bf 22}, 125 (1978).
%
\bibitem{Kashlinsky94} A. Kashlinsky, I. I.  Tkachev,  and  J. Frieman, Phys. Rev. Lett. {\bf 73}, 1582 (1994).
%
\bibitem{Langlois97} D. L. Langlois, Phys. Rev. D {\bf 55}, 7389 (1997).
%
\bibitem{Planckdf}  Planck Collaboration, Planck XIII, Astron. Astrophys. {\bf 561}, A97 (2014).
%
\bibitem{Atrio-Barandela13} F. Atrio-Barandela, Astron. Astrophys. {\bf 557}, A116 (2013).
%
\bibitem{Atrio-Barandela15} F. Atrio-Barandela, A. Kashlinsky, H. Ebeling, D. J. Fixen, and D. Kocevski, Astrophys. J. {\bf 810}, 143 (2015).
%
\bibitem{Ma2011} Y.-Z. Ma, C. Gordon, H. A. Feldman, Phys. Rev. D {\bf 83}, 103002 (2011).
%
\bibitem{Smoot92} G. F. Smoot $et ~ al.$ Astrophys. J Lett {\bf 396}, L1 (1992).
%
\bibitem{Fixen09} D. J. Fixen, Astrophys. J. {\bf 707}, 916 (2009).
%
\bibitem{Kogut93} A. Kogut  $et ~ al.$, Astrophys. J. {\bf 419}, 1 (1993).

\end{references}
\end{document}